\documentclass[letterpaper]{sfchem}
\usepackage{graphicx}

\begin{document}

\title{A Search for H$_2$ Outflow Signatures from Massive Star Formation Regions Containing Linearly Distributed Methanol Masers}

\author{James M. De Buizer} 
  \institute{Cerro Tololo Inter-American Observatory, Casilla 603, La Serena, Chile} 
\authorrunning{De Buizer}
\titlerunning{H$_2$ in Massive Star Forming Regions with Methanol Masers }

\maketitle 

\begin{abstract}
I present here the summary of results from a survey by De Buizer (2003) searching for
outflows using near-infrared imaging. Targets were 
massive young stellar objects associated with methanol masers in linear
distributions. Presently, it is a widely held belief that these methanol
masers are found in (and delineate) circumstellar accretion disks around
massive stars. A way to test the disk hypothesis is to
search for outflow signatures perpendicular to the methanol maser distributions.
The main objective of the survey was to obtain wide-field
near-infrared images of the sites of linearly distributed methanol masers
using a narrow-band 2.12 $\mu$m filter which is centered on the H$%
_{2}$ $v$=1--0 S(1) line. This line is a shock diagnostic that has been shown to
successfully trace CO outflows from young stellar objects. 
Twenty-eight sources in total were imaged with eighteen
sources displaying H$_2$ emission. Of these, only \textit{two} sources
showed emission found to be dominantly perpendicular to the methanol
maser distribution. These results seriously question the
hypothesis that methanol masers exist in circumstellar disks. 
 

\end{abstract}

\section{Introduction}
 
\begin{table*}[ht]
\begin{center}
\begin{minipage}{133mm}
\caption{List of targets observed in the De Buizer (2003) survey. The first column is the name of the target in galactic coordinates and the second column is the IRAS name of the target. The third and fourth columns give the source coordinates. The fifth column gives the methanol maser distribution position angle, and the sixth column describes the H$_2$ emission found in the target field. A `?' denotes that there may be some confusion associated with the nature or morphology of the H$_2$ emission.}
\begin{tabular}{lccccccc}
\hline
Target & IRAS Name & Right Ascension & Declination & Maser & H$_2$ \\
& & (J2000) & (J2000) & p.a. &  \\ 
\hline
G305.21+0.21   & 13079-6218 & 13 11 13.72 & -62 34 41.6 & 25\degr & parallel?    \\
G308.918+0.123 & 13395-6153 & 13 43 01.75 & -62 08 51.3 & 137\degr& parallel     \\
G309.92+0.48   & 13471-6120 & 13 50 41.76 & -61 35 10.1 & 30\degr & no detection \\
G312.11+0.26   & 14050-6056 & 14 08 49.30 & -61 13 26.0 & 166\degr& no detection \\
G313.77-0.86   & 14212-6131 & 14 25 01.62 & -61 44 58.1 & 135\degr& parallel     \\
G316.81-0.06   & 14416-5937 & 14 45 26.44 & -59 49 16.4 & 1\degr  & not outflow  \\
G318.95-0.20   &            & 15 00 55.40 & -58 58 53.0 & 151\degr& parallel     \\
G320.23-0.28   & 15061-5814 & 15 09 51.95 & -58 25 38.1 & 86\degr & parallel     \\
G321.031-0.484 & 15122-5801 & 15 15 51.64 & -58 11 17.4 & 0\degr  & parallel?    \\
G321.034-0.483 & 15122-5801 & 15 15 52.52 & -58 11 07.2 & 85\degr & parallel?    \\
G327.120+0.511 & 15437-5343 & 15 47 32.71 & -53 52 38.5 & 150\degr& no detection \\
G327.402+0.445 & 15454-5335 & 15 49 19.50 & -53 45 13.9 & 62\degr & no detection \\
G328.81+0.63   & 15520-5234 & 15 55 48.61 & -52 43 06.2 & 86\degr & parallel     \\
G331.132-0.244 & 16071-5142 & 16 10 59.74 & -51 50 22.7 & 90\degr & parallel     \\
G331.28-0.19   & 16076-5134 & 16 11 26.60 & -51 41 56.6 & 166\degr& perpendicular\\
G335.789+0.174 &            & 16 29 47.33 & -48 15 52.4 & 136\degr& parallel?    \\
G336.43-0.26   & 16303-4758 & 16 34 20.34 & -48 05 32.5 & 163\degr& no detection \\
G337.705-0.053 & 16348-4654 & 16 38 29.61 & -47 00 35.7 & 137\degr& no detection \\
G339.88-1.26   & 16484-4603 & 16 52 04.66 & -46 08 34.2 & 137\degr& ?            \\
G339.95-0.54   & 16455-4531 & 16 49 07.99 & -45 37 58.5 & 122\degr& no detection \\
G344.23-0.57   & 17006-4215 & 17 04 07.70 & -42 18 39.1 & 117\degr& no detection \\
G345.01+1.79   & 16533-4009 & 16 56 47.56 & -40 14 26.2 & 78\degr & parallel?    \\
G345.01+1.80   & 16533-4009 & 16 56 46.80 & -40 14 09.1 & 30\degr & parallel?    \\
G348.71-1.04   & 17167-3854 & 17 20 04.02 & -38 58 30.0 & 152\degr& not outflow  \\
G353.410-0.360 & 17271-3439 & 17 30 26.17 & -34 41 45.6 & 153\degr& not outflow  \\
G00.70-0.04    & 17441-2822 & 17 47 24.74 & -28 21 43.7 & 51\degr & no detection \\
G10.47-0.03    & 18056-1952 & 18 08 38.21 & -19 51 49.6 & 98\degr & no detection \\
G11.50-1.49    & 18134-1942 & 18 16 22.13 & -19 41 27.3 & 174\degr& perpendicular\\
\hline
\end{tabular}
\end{minipage}
\end{center}
\end{table*} 

\begin{figure}[ht]
\begin{center}
\resizebox{70mm}{!}{\includegraphics{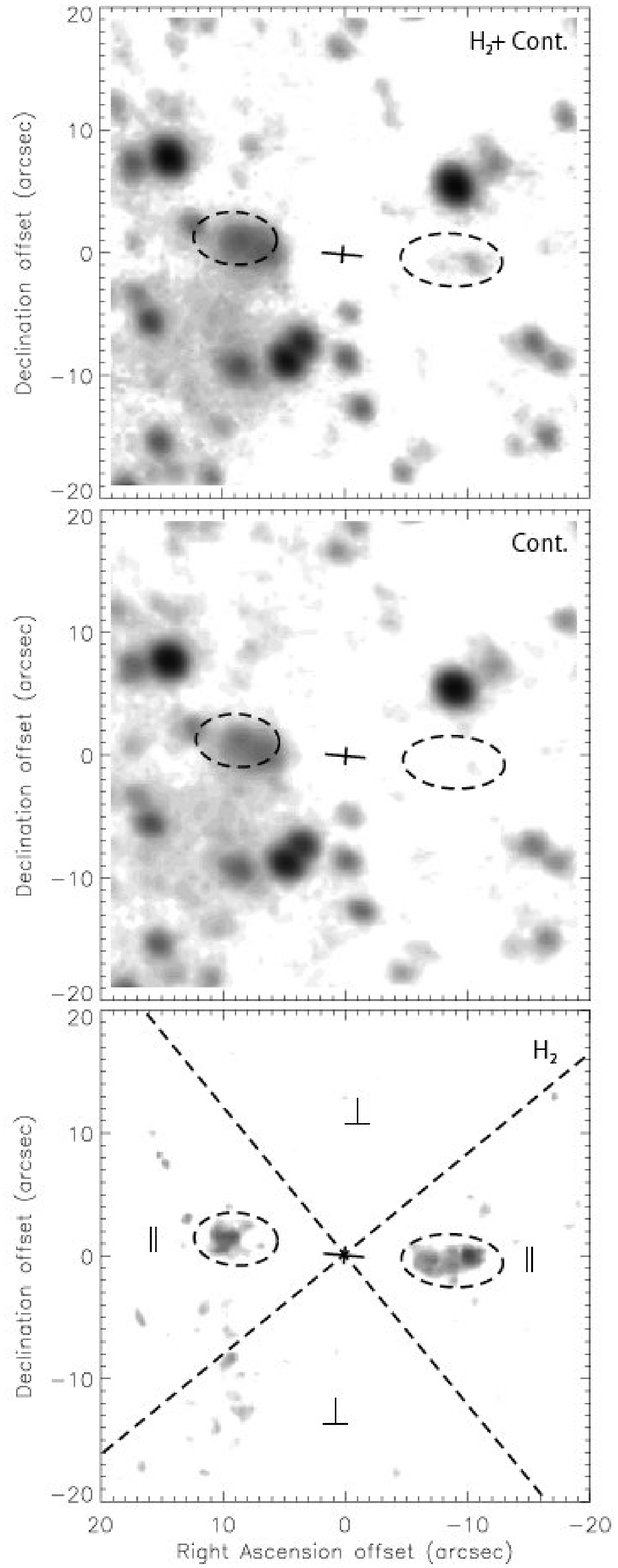}}
\caption{G320.23-0.28 (IRAS 15061-5814) H$_2$+continuum, continuum, and residual H$_2$ images. The cross represents the maser group location with the elongated axis showing the position angle of linear maser distribution. Dashed ellipses encompass areas of H$_2$ emission. Dashed lines in the H$_2$ image divide the frame into quadrants parallel and perpendicular to the maser position angle.
\label{fig-single}}
\end{center}
\end{figure}
 
Methanol maser emission has been well studied in the last decade and is thought to be a good indicator of recent massive star formation. These masers occur in spatially localized regions or {\it spots} and serve as powerful probes of the small-scale structure, dynamics, and physical conditions of the environments near forming stars. Radio observations by Norris et al. (1993) first showed that methanol (CH{$_3$}OH) maser spots are frequently distributed in linear structures, with projected dimensions typically spanning 2500 AU. Furthermore, some of these linearly distributed methanol masers exhibit a gradient in velocity along the spot distribution. Norris et al. (1993) argue that such a velocity gradient is indicative of orbital motion, and suggest that these methanol maser spots occur in, and directly delineate, rotating circumstellar disks. Several authors have written papers on radio studies of these methanol masers and their motions
(e.g. Norris et al. 1998; Phillips et al. 1998; Minier, Booth \& Conway 2000) trying to decipher from this limited data the mass of the central protostars and sizes of circumstellar disks in which these methanol masers
reside.

But do methanol masers really trace disks? More proof is needed than a line
of masers displaying (perhaps) orbital motion. Direct detection of these disks would certainly be conclusive, however observations have proven to be
problematic. Observations in the near-infrared suffer from the great extinction towards these regions.
It has also been argued that in the mid-infrared it is difficult to tell if one
is observing dust emission from a circumstellar accretion disk or dust
emission from the placental envelope (Vinkovi\'{c} et al. 2000). The far-infrared and sub-millimeter can probe the cool thermal emission from a dust disk with minimal contamination from the placental
envelope, however no far-infrared facility exists that has the sub-arcsecond resolution to observe disks around these sources. Angular resolution presently is a problem for the sub-millimeter as well, however the construction of SMA (Submillimeter Array) and ALMA (Atacama Large Millimeter Array) are underway, and are presently 5 to 10 years away from carrying out these types of observations.
Corroborative evidence that linearly distributed methanol masers exist in circumstellar disks, therefore, needs to come from something other than direct detection of the accretion disks.

Fortunately, there is an \textit{indirect} way of testing whether or not
linearly distributed methanol masers exist in accretion disks. According to
the standard model of accretion, during the phase of stellar formation where
the star is being fed by an accretion disk, it is also
undergoing mass loss through a bipolar outflow. This bipolar outflow is
perpendicular to the plane of the accretion disk, along the axis of
rotation. Therefore, one can search these sources of linearly distributed
methanol masers for evidence of outflow perpendicular to the methanol maser
position angle. Such evidence would create an extremely solid case for the hypothesis that these methanol masers exist in circumstellar
disks, without the need for their direct detection.

I present here the results of
the survey of De Buizer (2003), the purpose of which was to image the regions around sites of linearly distributed methanol masers in the near-infrared with a narrow-band 2.12 $\mu$m filter. This filter is centered on the H$_{2}$ $v$=1--0 S(1)
line, which is a shock indicator, and was convincingly shown to successfully
trace CO outflows from young stellar objects by Davis \& Eisl\"{o}ffel
(1995). Each site was also observed with a continuum filter which yields a
measure of the continuum-only flux from all the sources in the field. By
subtracting the continuum image from the image with the hydrogen
line, structures associated only with hydrogen
emission emanating from the locations of the methanol masers were found.

\section{Summary of Results}

The full results of this survey are presented in De Buizer (2003). The target list consisted of 28 sites compiled mostly from the articles
by Walsh et al. (1998), Phillips et al. (1998), and Norris et al. (1998). The coordinates for all of
these sources are shown in the Table 1. These sites all contain groups of
linearly distributed methanol masers, many with velocity gradients along
their distributions indicative of rotation, and therefore represent the best
circumstellar disk candidates.

Of the 28 maser targets observed by De Buizer (2003), H$_{2}$ emission was detected from 18
sites (64\%). The distribution of the H$_{2}$ emission from these sites
takes on three forms: 1) extended diffuse areas of H$_{2}$-dominated
emission; 2) individual knots or blobs of H$_{2}$-only emission that range
in number, sometimes cometary-shaped; and 3) some combination of extended H$%
_{2}$-dominated emission with knots of H$_{2}$-only sources.

The purpose of these observations was to try to confirm the hypothesis that
linearly distributed methanol masers exist in, and directly delineate,
circumstellar disks by searching for shock excited H$_2$ outflow signatures perpendicular to the
methanol maser position angle. Of the 18 maser locations observed to have H$_2$ emission, 
only \textit{two} displayed this morphology. The majority had H$_2$ emission found to be dominantly 
distributed within 45\degr\ of being \textit{parallel} to their
maser position angle (see Figure 1 for an example). H$_{2}$ emission can also be due to UV florescence, so for any one source in this survey, follow-up observations (in say, $^{12}$CO or HCO+) would be needed to confirm the outflow nature. However, regardless of the nature of the H$_2$ emission, these results are unequivocally contrary to the circumstellar disk hypothesis. If the H$_2$ emission is indeed from outflows, then a likely explanation is that at least some linearly distributed methanol masers may be directly associated with outflows. The methanol masers appear to be located coincident with a stellar source at the center of the outflows in most cases. Perhaps the masers trace the jets or outflow surfaces near the central (proto-)stellar source.

\end{document}